\documentclass{article}
\usepackage{authblk}
\usepackage{amsmath, amssymb}
\usepackage{graphicx}
\usepackage{hyperref}

\begin{document}

\title{Transition properties of Doubly Heavy Baryons}

\author{Kinjal Patel \thanks{kinjal1999patel@gmail.com}}

\author{Kaushal Thakkar\thanks{Corresponding Author: kaushal2physics@gmail.com}}

\affil{Department of Physics, Government College, Daman-396210, \\
U. T. of Dadra $\&$ Nagar Haveli and Daman $\&$ Diu, Veer Narmad South Gujarat University, Surat, India }

\maketitle

\begin{abstract}
In this study, we investigate the radiative and semileptonic decays of doubly heavy baryons (DHBs) within the framework of the Hypercentral Constituent Quark Model (HCQM). Our focus is on determining static and dynamic properties such as ground-state masses, magnetic moment, transition magnetic moment, radiative decay and heavy-to-heavy semileptonic decay rates, including their corresponding branching fractions. The ground-state masses are calculated using the six-dimensional hyper-radial Schr\"{o}dinger equation. The magnetic moments and transition magnetic moments for $J^P=\frac{1}{2}^+$ and $J^P=\frac{3}{2}^+$ baryons are also calculated. In addition, radiative decay widths are computed from the transition magnetic moment. We employed the Isgur-Wise function (IWF) to analyse the semileptonic decay widths of DHBs. The obtained results are compared with other theoretical predictions.\\
\end{abstract}

\noindent\textbf{Keywords:} Doubly heavy baryons, Radiative decays, Isgur-wise function, Semileptonic decays

\section{Introduction}\label{sec1}
All ground-state baryons with zero or one heavy quark have been experimentally well established \cite{pdg2024}. Research on baryons containing two or more charm or bottom quarks has gained interest in recent years. All DHBs with their quark contents and experimental statuses are shown in Table \ref{tab:Table1}. So far, only two doubly charmed baryons $\Xi_{cc}^{++}$ and $\Xi_{cc}^{+}$ have been observed experimentally \cite{pdg2024}. The first doubly charmed baryon $\Xi_{cc}^+$(3520) was reported by SELEX collaboration \cite{Mattson2002,Ocherashvili2005}. Before the experimental discovery of $\Xi_{cc}^{++}$ baryon, the theoretical study on the weak decays of DHBs have pointed out the most likely discovery of doubly charmed baryons $\Xi_{cc}^{++}$ via the two decay channels $\Xi_{cc}^{++} \rightarrow \Lambda^+_c K^-\pi^+\pi^-$ and $\Xi_{cc}^{++} \rightarrow \Xi_{cc}^{+}\pi^+$ \cite{Yu2018}. Later, the $\Xi_{cc}^{++}$ baryon was confirmed by LHCb Collaboration with mass and mean lifetime given as $m_{\Xi_{cc}^{++}} = 3621.55\pm0.23\pm0.30$ $MeV$ and $256^{+24}_{-22}\pm14$ $fs$, respectively \cite{Aaij2017,Aaij2018,Aaij2019,Aaij2020,Aaij2020a}. The spin-parity of both $\Xi_{cc}^+$ and $\Xi_{cc}^{++}$ are yet to be identified. A search for the doubly heavy $\Xi_{bc}^0$ baryon using its decay to the $D_0$p$K^-$ final state was performed using proton-proton collision data from the LHCb experiment, but no significant signal was found \cite{Aaij2020c}. LHCb reported the first search for the $\Omega_{bc}^0$ and a new search for the $\Xi_{bc}^0$ baryons in 2021. No significant excess was found for invariant predicted masses between 6.7 and 7.3 GeV/$c^2$\cite{Aaij2021}. A search for $\Xi_{cc}^+$(ccd) and $\Omega_{cc}^+$(ccs) was performed by LHCb Collaboration and only hints of signals were seen\cite{Aaij2020b,Aaij2021a,Aaij2021b}. The first search for the baryon using decays is reported by the LHCb experiment using a collision data sample which reported no significant signal \cite{Aaij2023}. Recently, a new mode of the doubly-charmed baryon decay, $\Xi_{cc}^{++} \rightarrow \Xi_{c}^{0}\pi^+ \pi^+$ observed in a data sample of pp collisions collected by the LHCb experiment \cite{Raaij2025}. The experimental as well as theoretical data for the masses, magnetic moments, semileptonic decay and other properties of singly heavy baryons are largely available. At the same time, there is no significant experimental data available for DHBs except for the $\Xi_{cc}^{++}$ baryon. Also, the intrinsic properties like magnetic moments remain unmeasured. The possibility to access the magnetic dipole moments and electric dipole moments of heavy and strange baryons at the LHC has been explored in recent years \cite{{Baryshevsky2016},{Burmistrov2016},{Fomin2017},{Baryshevsky2017},{Botella2017},{Bagli2017},{Henry},{Baryshevsky},{Scandale2016}}. To date, no experimental measurement of the magnetic moments of charm or beauty baryons has been successfully performed. The main obstacle is that these particles have short lifetimes, making it challenging to use standard magnetic moment measurement methods. Experimental measurements of the magnetic moments of DHBs would provide valuable information for low-energy QCD calculations. In the near future, the direct measurements of the magnetic moments of DHBs are uncertain. For this reason, any indirect prediction of the magnetic moments of DHBs could be crucial.\\

Pre-theoretical studies of their decay are important for experimental research on DHBs. Therefore, reliable theoretical predictions for these properties are required. Various properties of DHBs have been investigated via different theoretical approaches such as Quark model (QM) \cite{Roberts2008}, Quark-diquark model\cite{Farhadi2023}, Relativistic quark model (RQM)\cite{Ebert2002,Ebert2004,Ebert2005}, Non-relativistic quark model (NRQM) \cite{Brac1996,Gershtein2000,Ghalenovi2022,Song2024}, Light Front approach in diquark picture\cite{zhao2018}, QCD sum rule (QCDSR) \cite{Tousi2024,Zhang2008,Wang2010,Aliyev2022}, Heavy Diquark Effective Theory (HDiET)\cite{shi2020}, Born-Oppenheimer EFT \cite{Castella2021,Soto2020}, Bethe-Salpeter Equation\cite{Yu2019} and Lattice QCD (LQCD) \cite{Brown2014,Padmanath2019,Mathur2018}. The weak decays of DHBs have been studied extensively (See Ref. \cite{Lozano1994,Lozano1995,Li2017}). The magnetic moments of DHBs are investigated in Chiral Perturbation Theory (ChPT)\cite{{Li2017a},{Liu2018},{Blin2018},{Meng2017}}, Lattice QCD (LQCD) \cite{Can2014}, QCD light-cone sum rule (LCSR) \cite{Ozdem2023}. Radiative M1 transitions of heavy baryons are studied in modified bag model\cite{Bernotas2013}. Radiative Decays of the Spin-$\frac{3}{2}$ to Spin-$\frac{1}{2}$ Doubly heavy baryon are studied in light cone sum rules in Ref. \cite{Aliev2023}. In Ref. \cite{{Aliev2012},{Aliev2013},{Aliev2012a}}, the mixing angle, masses and residues of spin-$\frac{3}{2}$ and $\frac{1}{2}$ DHBs are studied within the QCD sum rules. Radiative decays of DHBs have been studied in non-relativistic potential model \cite{Albertus2010} and in relativistic three-quark model \cite{Branz2010}. The search for doubly charmed baryon $\Xi_{cc}^{++}$ with $J^P = \frac{3}{2}^+$ via it's electromagnetic transition is proposed in Ref. \cite{Cui2018}. The magnetic moments, M1 transition moments and radiative decay widths for all ground-state heavy baryons have been studied within the framework of extended bag model\cite{Simonis2018}.  In Ref. \cite{Hu2020}, the authors conducted a comprehensive analysis of the weak transition form factors for DHBs using the light front approach. The $b \rightarrow c$ decay form factors of DHBs in the QCD sum rules have been studied in Ref. \cite{Xing2021}. Semileptonic and nonleptonic four-body decays of doubly charmed baryons have been studied in Ref. \cite{Geng2018,Li2021} under flavour SU(3) symmetry. Ref. \cite{Wang2017a} has analysed the weak decays of DHBs under flavour SU(3) symmetry. The mass, production, decay and detection of DHBs have been discussed in Ref. \cite{Karliner2014}. The semileptonic decay of bottom baryons to charm baryons yields a significant source of knowledge regarding the internal structure of hadrons. The calculation of Isgur-Wise function (IWF) yields insights into the branching ratio, semileptonic decay width and the Cabibbo-Kobayashi-Maskawa (CKM) quark mixing matrix \cite{Isgur1990}.\\

This paper is organised as follows: In Section \ref{sec:2}, we discuss the theoretical framework for the quark model to compute the ground-state masses of DHBs. The magnetic moments, transition magnetic moments and radiative decay widths of DHBs are calculated in Section \ref{sec:3}. In Section \ref{sec:4}, we calculate the Isgur-Wise function and semileptonic decay width for heavy-to-heavy transition. The results are presented and discussed in Section \ref{sec:5}. The paper is summarised in Section \ref{sec:6}.

\begin{table}
\centering
    \caption{\label{tab:Table1}Doubly Heavy Baryons}
    \begin{tabular}{ccc}
    \hline
    Baryon & quark & Experimental \\
    &content&status \cite{pdg2024}\\
    \hline
    $\Xi_{bb}^-$ & bbu  & - \\
    $\Xi_{bb}^0$ & bbd & - \\
    $\Xi_{cc}^{++}$ & ccu & *** \\
    $\Xi_{cc}^+$ & ccd & * \\
    $\Xi_{bc}^+$ & bcu & - \\
    $\Xi_{bc}^0$ & bcd & -  \\
    $\Omega_{bb}^-$ & bbs & - \\
    $\Omega_{bc}^0$ & bcs & - \\
    \hline
    \end{tabular}
\end{table}

\section{Theoretical Framework}\label{sec:2}

We adopted the Hypercentral constituent quark model (HCQM) to study DHBs. We consider a doubly heavy baryon to be a bound state of two heavy quarks and one light quark. Conventional quark models vary in their assumptions, but they share a basic structure and certain fundamental traits, such as confinement and asymptotic freedom, with the remaining aspects being constructed through appropriate assumptions. The main differences between the framework of this study and the Quark model adopted in Ref. \cite{Roberts2008} are expressed as follow:\\
\begin{enumerate}
    \item The masses of the light quarks ($u$ and $d$) were the same ($m_u$ = $m_d$) in the Quark model, whereas in this study, we used unequal quark masses ($m_u \neq m_d$) in HCQM.
    \item In Quark model, the Schr\"{o}dinger equation is solved in three-dimensional space, whereas we have solved the Schr\"{o}dinger equation in six-dimensional space.
    \item The confinement potential used in Isgur and Karl quark model is given by harmonic oscillator plus constant potential, while the confinement potential used in HCQM is given by linear plus hyper Coulomb potential.
    \item The main difference here is that the potential $V(x)$ is not purely a two-body interaction but it contains three-body effects also. The three-body effects are desirable in the study of hadrons since the non-abelian nature of QCD leads to gluon-gluon couplings which produce three-body forces as given in Ref. \cite{Santopinto1998}.
\end{enumerate}

The dynamics of the three quarks can be described using Jacobi coordinates. The hyperspherical coordinates: hyper radius and hyper angle are defined in terms of Jacobi coordinates\cite{Thakkar2017,Thakkar2020}.
\begin{equation}\label{eq:1}
\rho = \frac{1}{\sqrt{2}}(\mathbf{r}_1 - \mathbf{r}_2)
\end{equation}
\begin{equation}\label{eq:2}
\lambda = \frac{{m_1 \mathbf{r}_1 + m_2 \mathbf{r}_2 - (m_1 + m_2) \mathbf{r}_3}}{{\sqrt{m_1^2 + m_2^2 + (m_1 + m_2)^2}}}
\end{equation}
Such that,
\begin{equation}\label{eq:2a}
m_{\rho} = \frac{2m_1m_2}{m_1+m_2}
\end{equation}
\begin{equation}\label{eq:2b}
m_{\lambda} = \frac{2m_3(m_1^2+m_2^2+m_1m_2)}{(m_1+m_2)(m_1+m_2+m_3)}
\end{equation}
where $m_1$, $m_2$, and $m_3$ are the constituent quark masses. The hyperspherical coordinates are given by the angles $\Omega_\rho=(\theta_\rho,\phi_\rho)$ and $\Omega_\lambda=(\theta_\lambda,\phi_\lambda)$. The hyper-radius $x$ and hyper-angle $\xi$ are defined as
\begin{equation}\label{3}
x=\sqrt{\rho^2+\lambda^2};\xi=\arctan\left(\frac{\rho}{\lambda}\right)
\end{equation}
 Using hyperspherical coordinates, the kinetic energy operator $P_x^2/2m$ of the three-body system can be written as
\begin{equation}\label{4}
\frac{P_x^2}{2m} = -\frac{\hbar^2}{2m} \left(\frac{\partial^2}{\partial x^2} + \frac{5}{x} \frac{\partial}{\partial x} + \frac{L^2(\Omega_\rho,\Omega_\lambda,\xi)}{x^2}\right)
\end{equation}
where $m= \frac{2m_\rho m_\lambda}{m_\rho+m_\lambda}$ denotes reduced mass. $L^2(\Omega_\rho,\Omega_\lambda,\xi)$ is the quadratic Casimir operator of the six-dimensional rotational group $O(6)$ and its eigenfunctions are the hyperspherical harmonics,\\ $Y_{[\gamma]l_\rho l_\lambda}(\Omega_\rho,\Omega_\lambda,\xi)$ which satisfies the eigenvalue relation
\begin{equation}\label{5}
L^2Y_{[\gamma]l_{\rho}l_{\lambda}}(\Omega_{\rho},\Omega_{\lambda},\xi)=\gamma(\gamma+4)Y_{[\gamma]l_{\rho}l_{\lambda}}(\Omega_{\rho},\Omega_{\lambda},\xi)
\end{equation}
where, $l_\rho$ and $l_\lambda$ are the angular momenta associated with the $\rho$ and $\lambda$ variables respectively. The model Hamiltonian for baryons can be expressed as
\begin{equation}\label{eq:6}
H = \frac{P_\rho^2}{2m_\rho}+\frac{P_\lambda^2}{2m_\lambda}+V(\rho,\lambda) = \frac{P_x^2}{2m} + V(x)
\end{equation}
Here, the potential $V(x)$ is not purely a two-body interaction, but also contains three-body effects. The six-dimensional hyper-radial Schr\"{o}dinger equation can be written as
\begin{eqnarray}\label{eq:7}
\left[ \frac{d^2}{dx^2} + \frac{5}{x}\frac{d}{dx} - \frac{\gamma(\gamma + 4)}{x^2} \right] \psi_{\nu\gamma}(x) = -2m[E - V(x)]\psi_{\nu\gamma}(x)
\end{eqnarray}
where $\psi_{\nu\gamma}$ is the hyper-radial wave function. The potential is assumed to depend only on the hyper radius and hence is a three-body potential because the hyper radius depends only on the coordinates of all three quarks. The hyper-Coulomb plus linear potential is given as
\begin{equation}\label{eq:8}
V(x) = \frac{\tau}{x} + \beta x + V_0 + V_{spin}
\end{equation}
    where $\tau$ = $-$$\frac{2}{3}\alpha_s$ is the hyper-Coulomb strength and the values of $\beta$ and $V_0$ are fixed to obtain the ground-state masses. $V_{spin}$ is the spin-dependent part, given as \cite{Garcilazo2007}\\
\begin{equation}\label{eq:9}
V_{spin}(x) = -\frac{A}{4} \alpha_s \vec{\lambda}_i \cdot \vec{\lambda}_j \frac{e^{-x/x_0}}{x {x_0}^2} \sum_{i<j} \frac{\vec{\sigma}_i \cdot \vec{\sigma}_j}{6 m_i m_j}
\end{equation}
Here, the parameter $A$ and regularization parameter $x_0$ are considered as the hyperfine parameters of the model. $\lambda_{i,j}$ are the SU(3) colour matrices, and $\sigma_{i,j}$ are spin Pauli matrices, $m_{i,j}$ are the constituent masses of the two interacting quarks. The parameter $\alpha_s$ corresponds to the strong running coupling constant, which is given as
\begin{equation}\label{10}
\alpha_s = \frac{\alpha_s(\mu_0)}{1+(\frac{33-2n_f}{12\pi})\alpha_s(\mu_0)ln(\frac{m_1+m_2+m_3}{\mu_0})}
\end{equation}
We factor out the hyper angular part of the three-quark wave function is given by hyperspherical harmonics. The hyperradial part of the wave function is evaluated by solving the Schr\"{o}dinger equation. The hyper-Coulomb trial radial wave function is given by \cite{Santopinto1998,Ferraris1995,Thakkar2011}

\begin{eqnarray}\label{eq:11}
\psi_{\nu\gamma}=\left[\frac{(\nu - \gamma)! (2g)^6}{(2\nu + 5)(\nu + \gamma + 4)!}\right]^{\frac{1}{2}} (2gx)^\gamma \times e^{-gx} L^{2\gamma + 4}_{\nu-\gamma} (2gx)
\end{eqnarray}

    Here, $\gamma$ is the hyperangular quantum number and $\nu$ denotes the number of nodes of the spatial three-quark wave function. $L^{2\gamma + 4}_{\nu-\gamma}(2gx)$ is the associated Laguerre polynomial. The wave function parameter $g$ and energy eigenvalues are obtained by applying the virial theorem. The masses of the ground-state DHBs are calculated by summing the model quark masses (see Table \ref{tab:Table2}), kinetic energy and potential energy.
\begin{equation}\label{eq:12}
M_B = m_1 + m_2 + m_3 + \langle H \rangle
\end{equation}
The computed ground-state masses of DHBs and percentage errors are listed and compared in Table \ref{tab:Table3}.


\begin{table}
\centering
\caption{\label{tab:Table2}Quark mass parameters (in GeV) and constants used in the calculations.}
    \begin{tabular}{cc}
    \hline
    Parameter & Value\\
    \hline
    ${m_{u}}$ & 0.33\\
    ${m_{d}}$ & 0.35\\
    ${m_{s}}$ & 0.50\\
    ${m_{c}}$ & 1.55\\
    ${m_{b}}$ & 4.95\\
    $\alpha_s(\mu_{0}$=1 GeV) & 0.6\\
    $\beta$ & 0.14\\
    ${V_0}$ & -0.818\\
    $x_0$ & 1\\
    \hline
    \end{tabular}
\end{table}

\begin{table*}
\begin{center}
    \caption{{\label{tab:Table3} Ground state masses of doubly heavy baryons in GeV and percentage error}}
    \tiny
    \begin{tabular}{ccccccccccc}
    \hline
    Baryons	&	Our	&	\cite{Brown2014}	&	\cite{Ebert2005}	&	 \cite{Gershtein2000}	&	\cite{Rahmani2020}	&	\cite{Tousi2024}	&	 \cite{Karliner2014}	&	 \cite{Mathur2018}	&	$\%$ Error \cite{Brown2014}	 &	$\%$ Error \cite{Mathur2018}	\\
    \hline
    $\Xi_{bb}^0$	&	10.242	&	10.143	&	10.202	&	10.093	&	10.215	 &	$9.97\pm0.19$	&	$10.162\pm0.012$	&	-	&	0.98	&	-	\\
    $\Xi_{bb}^-$	&	10.246	&	10.143	&	10.202	&	10.093	&	10.215	 &	$9.97\pm0.19$	&	$10.162\pm0.012$	&	-	&	1.02	&	-	\\
    $\Xi_{bc}^+$	&	6.855	&	6.943	&	6.933	&	6.82	&	6.805	 &	$6.73^{+0.14}_{-0.13}$	&	$6.914\pm0.013$	&	6.945	&	1.27	&	 1.30	\\
    $\Xi_{bc}^0$	&	6.861	&	6.943	&	6.933	&	6.82	&	6.805	 &	$6.73^{+0.14}_{-0.13}$	&	$6.914\pm0.013$	&	-	&	1.19	&	-	 \\
    $\Xi_{cc}^{++}$	&	3.457	&	3.61	&	3.62	&	3.478	&	3.396	 &	$3.69\pm0.10$	&	$3.627\pm0.012$	&	-	&	4.25	&	-	\\
    $\Xi_{cc}^+$	&	3.464	&	3.61	&	3.62	&	3.478	&	3.396	 &	$3.69\pm0.10$	&	$3.627\pm0.012$	&	-	&	4.05	&	-	\\
    $\Omega_{bb}^-$	&	10.309	&	10.273	&	10.359	&	10.18	&	10.364	 &	$9.98\pm0.18$	&	-	&	-	&	0.35	&	-	\\
    $\Omega_{bc}^0$	&	6.932	&	6.998	&	7.088	&	6.910	&	6.958	 &	$6.77^{+0.13}_{-0.12}$	&	-	&	6.994	&	0.94	&	0.89	\\
    $\Omega_{cc}^+$	&	3.547	&	3.738	&	3.778	&	3.590	&	3.552	 &	$3.70\pm0.09$	&	-	&	-	&	5.09	&	-	\\
    $\Xi_{bb}^{0*}$	&	10.262	&	10.178	&	10.237	&	10.133	&	10.227	 &	-	&	$10.184\pm0.012$	&	-	&	0.82	&	-	\\
    $\Xi_{bb}^{-*}$	&	10.266	&	10.178	&	10.237	&	10.133	&	10.227	 &	-	&	$10.184\pm0.012$	&	-	&	0.86	&	-	\\
    $\Xi_{bc}^{*+}$	&	6.897	&	6.985	&	6.980	&	6.9	&	6.83	&	 -	&	$6.969\pm0.014$	&	6.989	&	1.25	&	1.31	\\
    $\Xi_{bc}^{0*}$	& 6.902 &	6.985	&	6.980	&	6.9	&	6.83	&	-	 &	$6.969\pm0.014$	&	-	&	1.18	&	-	\\
    $\Xi_{cc}^{++*}$ & 3.539	&	3.692	&	3.727	&	3.61	&	3.434	 &	-	&	$3.690\pm0.012$	&	-	&	4.14	&	-	\\
    $\Xi_{cc}^{+*}$	& 3.545	&	3.692	&	3.727	&	3.61	&	3.434	&	 -	&	$3.690\pm0.012$	&	-	&	3.98	&	-	\\
    $\Omega_{bb}^{-*}$ & 10.328	&	10.308	&	10.389	&	10.200	&	10.372	 &	-	&	-	&	-	&	0.19	&	-	\\
    $\Omega_{bc}^{0*}$ & 6.971	&	7.059	&	7.130	&	6.99	&	6.975	 &	-	&	-	&	7.056	&	1.24	&	1.20	\\
    $\Omega_{cc}^{+*}$ & 3.619	&	3.822	&	3.872	&	3.69	&	3.578	 &	-	&	-	&	-	&	5.31	&	-	\\
    \hline
    \end{tabular}
\end{center}
\end{table*}

\section{Magnetic Moment and Radiative decay}\label{sec:3}
\textbf{\subsection{Effective quark masses and magnetic moment for doubly heavy baryons}}
The electromagnetic properties of baryons are an important source of information regarding their internal structure. The magnetic moments of baryons are obtained in terms of the spin-flavour wave function of the constituent quarks as \cite{majethiya2008}:
\begin{equation}\label{eq:13}
\mu_B = \Sigma_i \langle \phi_{sf}|\mu_i|\phi_{sf}\rangle
\end{equation}
where
\begin{equation}\label{eq:14}
\mu_i = \frac{e_i\sigma_i}{2m_i^{eff}}
\end{equation}
where $i$ = u,d,s,c,b; $e_i$ and $\sigma_i$ represent the charge and spin of the constituting quarks of the baryonic state, respectively; $|\phi_{sf}\rangle$ represents the spin-flavour wave function of the respective baryonic state. The expressions for the magnetic moments of $J^P=\frac{1}{2}^+$ and $J^P=\frac{3}{2}^+$ for DHBs are given in Table \ref{tab:table5}. Here, $m_i$ the mass of $i^{th}$ quark in the three-body baryon, is taken as an effective mass of the constituting quarks, as their motions are governed by the three-body force described by the Hamiltonian in Eq. (\ref{eq:6}). The baryon mass of the quarks may get modified due to their binding interactions with the other two quarks. We account for this bound state effect by replacing the mass parameter $m_i$ in Eq. (\ref{eq:14}) by defining an effective mass to the bound quarks, $m_{i}^{eff}$ is given as \cite{Thakkar2011}

\begin{equation}\label{eq:15}
m_i^{eff} = m_i \left( 1+\frac{\langle H \rangle}{\sum_i m_i}\right)
\end{equation}

such that $M_B = \sum_{i=1}^3 m_i^{eff} $where $\langle H \rangle$ = E + $\langle V(x) \rangle$. The calculated magnetic moments for DHBs are listed and compared with other theoretical models in Table \ref{tab:table6}.\\

\textbf{\subsection{Transition magnetic moment and radiative decay width}}
The transition magnetic moment for $\frac{3}{2}^+ \rightarrow \frac{1}{2}^+$ can be expressed as \cite{Thakkar2011}
\begin{equation}\label{eq:16}
\mu_{\frac{3}{2}^+ \rightarrow \frac{1}{2}^+} = \sum_i \left\langle \phi_{sf}^{\frac{3}{2}^+} | \mu_i\sigma_i| \phi_{sf}^{\frac{1}{2}^+} \right\rangle
\end{equation}
$\langle \phi_{sf}^{\frac{3}{2}^+} |$ represents the spin-flavour wave function of the quark composition for the respective baryons with $J^P=\frac{3}{2}^+$ while $|\phi_{sf}^{\frac{1}{2}^+}\rangle$ represent the spin-flavour wave function of the quark composition for the baryons $J^P=\frac{1}{2}^+$. To compute the transition magnetic moment ($\mu_{\frac{3}{2}^+ \rightarrow \frac{1}{2}^+}$), we take the geometric mean of the effective quark masses of the constituent quarks of initial and final state baryons,
\begin{equation}\label{eq:17}
m_i^{eff} = \sqrt{m_{i{B^*}}^{eff}m_{iB}^{eff}}
\end{equation}
 Here, $m_{i{B^*}}^{eff}$ and $m_{iB}^{eff}$ are the effective masses of the quarks constituting the baryonic states $B^*$ and $B$, respectively. Considering the geometric mean of the effective quark masses of the constituting quarks and the spin-flavour wave functions of the baryonic states, the transition magnetic moments are computed using Eq. (\ref{eq:16}). The expressions for the transition magnetic moments and the obtained transition magnetic moments of the DHBs are listed in Table \ref{tab:table7}. We can see that the results are in accordance with other theoretical predictions.

The radiative decay width can be expressed in terms of the radiative transition magnetic moment and photon momentum ($k$) as \cite{Bernotas2013,Wagner2000}
\begin{equation}\label{eq:18}
 \Gamma = \frac{\alpha k^3}{M_P^2} \frac{2}{2J+1}\frac{M_B}{M_{B^*}} \mu^2(B^* \rightarrow B\gamma)
\end{equation}
where, $\mu^2(B^* \rightarrow B\gamma)$ is the square of the transition magnetic moment, $\alpha=\frac{1}{137}$ and $M_P$ is the mass of the proton = 0.938 GeV. $J$ and $M_{B^*}$ are the total angular momentum and mass of the decaying baryon, respectively. $M_B$ is the final state baryon mass. $k$ is the photon momentum in the center-of-mass system of the decaying baryon, given by
\begin{equation}\label{eq:19}
k = \frac{M^2_{B^*} - M^2_B}{2M_B}
\end{equation}
Here, we ignore $E2$ amplitudes because of the spherical symmetry of the S-wave baryon spatial wave function and the radiative width of the decay $B^* \rightarrow B\gamma$ has the form of Eq. (\ref{eq:18}). The calculated radiative decay widths are presented and compared in Table \ref{tab:table8}.\\

\begin{table*}\caption{Expressions of magnetic moments for doubly heavy baryons}
\begin{center}
\centering
    \label{tab:table5}
    \begin{tabular}{ccc}
        \hline
        &\multicolumn{2}{c}{\underline{Magnetic Moment Expressions}} \\
        Baryon &  $J^P=\frac{1}{2}^+$  &  $J^P=\frac{3}{2}^+$ \\
        \hline
        $\Xi_{cc}^{++}$ &  $\frac{4}{3} \mu_c - \frac{1}{3} \mu_u$  & $2\mu_c + \mu_u$ \\
        $\Xi_{cc}^{+}$  & $\frac{4}{3} \mu_c - \frac{1}{3} \mu_d$  & $2\mu_c + \mu_d$\\
        $\Xi_{bb}^{0}$ & $\frac{4}{3} \mu_b - \frac{1}{3} \mu_u$ & $2\mu_b + \mu_u$\\
        $\Xi_{bb}^{-}$ &  $\frac{4}{3} \mu_b - \frac{1}{3} \mu_d$ & $2\mu_b + \mu_d$\\
        $\Xi_{bc}^{+}$ & $\frac{2}{3} \mu_b + \frac{2}{3} \mu_c - \frac{1}{3} \mu_u$  & $\mu_b + \mu_c + \mu_u$\\
        $\Xi_{bc}^{0}$ & $\frac{2}{3} \mu_b + \frac{2}{3} \mu_c - \frac{1}{3} \mu_d$ & $\mu_b + \mu_c + \mu_d$\\
        $\Omega_{bb}^{-}$  & $\frac{4}{3} \mu_b - \frac{1}{3} \mu_s$ & $2\mu_b + \mu_s$\\
        $\Omega_{bc}^{0}$ & $\frac{2}{3} \mu_b + \frac{2}{3} \mu_c - \frac{1}{3} \mu_s$ & $\mu_b + \mu_c + \mu_s$\\
        $\Omega_{cc}^{+}$ &  $\frac{4}{3} \mu_c - \frac{1}{3} \mu_s$ & $2\mu_c + \mu_s$\\
        \hline
    \end{tabular}
    \end{center}
\end{table*}

\begin{table*}
\begin{center}
\centering
\caption{\label{tab:table5} Magnetic Moment of $J^P=\frac{1}{2}^+$ doubly heavy baryons in $\mu_N$}
\tiny
\begin{tabular}{ccccccccccc}
\hline
    Baryon 	&	Our	&	 \cite{Gadaria2016}	&	 \cite{shah2021}	&	 \cite{Hazra2021}	&	\cite{Ozdem2019}	&	\cite{Ozdem2023}	&	 \cite{Can2014}	&	 \cite{Li2017a}	&	 \cite{Bahtiyar2022}	&	 \cite{Sharma2010}	\\
    \hline
    $\Xi_{bb}^{0}$	&	-0.715	&	-0.89	&	-0.663	&	$-0.6699\pm0.0006$	 &	$-0.51\pm0.09$	 &	-	&	-	&	-0.84	&	-	 &	-	\\
    $\Xi_{bb}^{-}$	&	0.214	&	0.32	&	0.196	&	$0.2108\pm0.0003$	 &	$0.28\pm0.04$	 &	-	&	-	&	0.26	&	-	 &	-	\\
    $\Xi_{bc}^{+}$	&	-0.403	&	-0.52	&	-0.304	&	$-0.06202\pm0.00001$	 &	-	 &	$-0.50^{+0.14}_{-0.12}$	&	-	&	-0.54	&	-	 &	-	\\
    $\Xi_{bc}^{0}$	&	0.524	&	0.63	&	0.527	&	$-0.06202\pm0.00001$	 &	-	 &	$0.39^{+0.06}_{-0.05}$	&	-	&	0.56	&	-	 &	-	\\
    $\Xi_{cc}^{++}$	&	-0.093	&	-0.169	&	0.031	&	$-0.1046\pm0.0021$	 &	$-0.23\pm0.05$	 &	-	&	$0.425\pm0.029$	&	-0.25	&	 $0.430\pm0.019$	 &	0.006	 \\
    $\Xi_{cc}^{+}$	&	0.832	&	0.853	&	0.784	&	$0.8148\pm0.0018$	 &	$0.43\pm0.09$	 &	-	&	-	&	0.85	&	-	 &	0.84	\\
    $\Omega_{bb}^{-}$	&	0.125	&	0.16	&	0.108	&	 $0.1135\pm0.0008$	&	$0.42\pm0.05$	&	-	&	-	&	0.19	&	-	 &	-	\\
    $\Omega_{bc}^{0}$	&	0.439	&	0.49	&	-	&	$-0.06202\pm0.00001$	 &	-	 &	$0.38^{+0.05}_{-0.04}$	&	-	&	0.49	&	-	 &	-	\\
    $\Omega_{cc}^{+}$	&	0.757	&	0.74	&	0.692	&	 $0.7109\pm0.0017$	&	$0.39\pm0.09$	&	-	&	$0.413\pm0.024$	&	0.78	 &	$0.433\pm0.039$	&	0.697	 \\
    \hline
    \end{tabular}
\end{center}
\end{table*}

\begin{table*}
\begin{center}
\centering
\caption{\label{tab:table6} Magnetic Moment of $J^P=\frac{3}{2}^+$ doubly heavy baryons in $\mu_N$}
\begin{tabular}{ccccccccc}
    \hline
    Baryon 	&	Our	&	 \cite{Gadaria2016}	&	 \cite{shah2021}	&	 \cite{Hazra2021}	&	\cite{Ozdem2020}	&	Set-1 \cite{Meng2017}	&	 Set-2 \cite{Meng2017}	&	 \cite{Sharma2010}	\\
    \hline
    $\Xi_{bb}^{0*}$	 &	1.763	&	2.30	&	-1.607	&	$1.5897\pm0.0016$	 &	2.30	&	-1.33	&	-1.38	&	-	\\
    $\Xi_{bb}^{-*}$	 &	-1.018	&	-1.32	&	-1.737	&	$-0.9809\pm0.0008$	 &	-1.39	&	2.83	&	2.87	&	-	\\
    $\Xi_{bc}^{*+}$	 &	2.213	&	2.68	&	2.107	&	$2.0131\pm0.0020$	 &	2.63	&	3.22	&	3.27	&	-	\\
    $\Xi_{bc}^{0*}$	 &	-0.549	&	-0.76	&	-0.448	&	$-0.5315\pm0.0012$	 &	-0.96	&	-0.84	&	-0.89	&	-	\\
    $\Xi_{cc}^{++*}$	 &	2.619	&	2.72	&	2.218	&	 $2.4344\pm0.0033$	 &	2.94	&	3.51	&	3.63	&	2.66	\\
    $\Xi_{cc}^{+*}$	 &	-0.084	&	-0.23	&	0.068	&	$-0.0846\pm0.0025$	 &	-0.67	&	-0.27	&	-0.37	&	-0.47	\\
    $\Omega_{bb}^{-*}$	&	-0.757	&	-0.86	&	-1.239	&	 $-0.6999\pm0.0017$	&	-1.56	&	-1.54	&	-1.55	&	-	\\
    $\Omega_{bc}^{0*}$	 &	-0.286	&	-0.32	&	-	&	$-0.2552\pm0.0016$	 &	-1.11	&	-1.09	&	-1.10	&	-	\\
    $\Omega_{cc}^{+*}$	&	0.181	&	0.16	&	0.285	&	 $0.1871\pm0.0026$	&	-0.52	&	-0.64	&	-0.65	&	0.14	\\
    \hline
    \end{tabular}
\end{center}
\end{table*}

\begin{table*}
\begin{center}
\centering
\caption{\label{tab:table7} Transition magnetic moments in $\mu_N$}
\begin{tabular}{ccccccccc}
    \hline
    Transition 	&	 Expression 	&	 our 	&	 \cite{shah2021} 	&	 \cite{Bernotas2013} 	&	  \cite{Simonis2018} 	&	 \cite{Li2018} 	&	 \cite{Sharma2010}	&	 \cite{Aliev2023}	\\
    \hline																	 \\
    $\Xi_{bb}^{0*} \rightarrow \Xi_{bb}^{0}$ 	&	 $\frac{2\sqrt{2}}{3}(\mu_b-\mu_u)$ 	&	 $-1.842$ 	&	 $-1.69$ 	&	 $-1.039$ 	&	 $-1.45$ 	&	 $-1.81$ 	&	 -	&	 $1.78\pm0.300$	\\
    $\Xi_{bb}^{-*} \rightarrow \Xi_{bb}^{-}$ 	&	 $\frac{2\sqrt{2}}{3}(\mu_b-\mu_d)$ 	&	0.782	&	0.73	&	0.428	&	 0.643	&	0.81	&	-	&	 $0.82\pm0.131$	 \\
    $\Xi_{bc}^{+*} \rightarrow \Xi_{bc}^{+}$ 	&	 $\frac{\sqrt{2}}{3}(\mu_b+\mu_c-2\mu_u)$ 	&	 $-1.615$ 	&	 $-1.39$ 	 &	0.695	&	 $-1.37$ 	&	 $-1.61$	&	 -	&	 $1.25\pm0.156$	\\
    $\Xi_{bc}^{0*} \rightarrow \Xi_{bc}^{0}$ 	&	 $\frac{\sqrt{2}}{3}(\mu_b+\mu_c-2\mu_d)$ 	&	0.998	&	0.94	&	 $-0.747$ 	&	0.879	&	1.02	&	-	&	 $0.77\pm0.104$	\\
    $\Xi_{cc}^{++*}  \rightarrow\Xi_{cc}^{++}$ 	&	 $ \frac{4}{3\sqrt{2}}(\mu_c-\mu_u)$ 	&	 $-1.379$ 	&	 $-1.01$ 	&	 $-0.787$ 	&	 $-1.21$ 	&	 $-2.35$ 	&	 1.33	 &	$1.03\pm0.138$	 \\
    $\Xi_{cc}^{+*} \rightarrow \Xi_{cc}^{+}$ 	&	 $\frac{4}{3\sqrt{2}}(\mu_c-\mu_d)$ 	&	1.204	&	1.048	&	0.945	&	 1.07	&	1.55	&	-1.41	&	 $0.96\pm0.158$	\\
    $\Omega_{bb}^{-*} \rightarrow \Omega_{bb}^{-}$	&	 $ \frac{2\sqrt{2}}{3}(\mu_b-\mu_s)$ 	&	0.534	&	0.48	&	0.307	&	 0.478	&	0.48	&	 -	&	-	\\
    $\Omega_{bc}^{0*} \rightarrow \Omega_{bc}^{0}$	&	 $\frac{\sqrt{2}}{3}(\mu_b+\mu_c-2\mu_s)$  	&	0.755	&	0.71	&	 $-0.624$ 	&	0.688	&	0.69	&	-	&	 -	 \\
    $\Omega_{cc}^{+*} \rightarrow \Omega_{cc}^{+}$ 	&	 $\frac{4}{3\sqrt{2}}(\mu_c-\mu_s)$ 	&	0.974	&	0.96	&	0.789	&	 0.869	&	1.54	&	-0.89	&	-	 \\
    \hline
    \end{tabular}
    \end{center}
\end{table*}

\begin{table}
\begin{center}
\centering
\caption{\label{tab:table8} Radiative decay width of doubly heavy baryons in keV}
\small
\begin{tabular}{ccccccccc}
    \hline
    Transition 	&	 our 	&	 \cite{Bernotas2013} 	&	 \cite{Farhadi2023} 	 &	 \cite{Hazra2021} 	&	\cite{Branz2010} 	&	 \cite{Lu2017} 	&	 \cite{Aliev2023}	 \\
    \hline															
    $\Gamma(\Xi_{bb}^{0*} \rightarrow \Xi_{bb}^{0}\gamma)$ 	&	0.104	&	 0.126	&	 $0.40\pm0.044$ 	&	 $0.5509\pm0.023$ 	&	 $0.31\pm0.06$ 	 &	0.98	&	 $0.58\pm0.188$	 \\
    $\Gamma(\Xi_{bb}^{-*} \rightarrow \Xi_{bb}^{-}\gamma)$ 	&	0.018	&	 0.022	&	- 	&	$0.102\pm0.005$ 	&	 $0.059\pm0.014$ 	&	0.28	 &	$0.12\pm0.038$	 \\
    $\Gamma(\Xi_{bc}^{+*} \rightarrow \Xi_{bc}^{+} \gamma)$ 	&	0.812	&	 0.533	&	 $0.205\pm0.009$ 	&	 $0.381\pm0.017$ 	&	 $0.49\pm0.09$ 	 &	-	&	 $0.48\pm0.119$	 \\
    $\Gamma(\Xi_{bc}^{0*} \rightarrow \Xi_{bc}^{0}\gamma)$ 	&	0.304	&	 0.612	&	 -	&	 $0.321\pm0.014$ 	&	 $0.24\pm0.04$ 	&	-	&	 $0.18\pm0.049$	\\
    $\Gamma(\Xi_{cc}^{++*}  \rightarrow\Xi_{cc}^{++}\gamma)$ 	&	4.149	&	 1.43	&	 $2.22\pm0.098$ 	&	 $2.37\pm0.05$ 	&	 $23.46\pm3.33$ 	 &	7.21	&	 $2.36\pm0.622$	\\
    $\Gamma(\Xi_{cc}^{+*} \rightarrow \Xi_{cc}^{+}\gamma)$ 	&	3.059	&	2.08	 &	 -	&	 $1.98\pm0.04$ 	&	 $28.79\pm2.51$ 	&	3.9	&	 $2.07\pm0.666$	\\
    $\Gamma(\Omega_{bb}^{-*} \rightarrow \Omega_{bb}^{-}\gamma)$ 	&	0.008	 &	0.011	&	 $0.051\pm0.018$ 	&	 $0.0426\pm0.0018$ 	&	 $0.0226\pm0.0045$ 	&	0.04	 &	-	 \\
    $\Gamma(\Omega_{bc}^{0*} \rightarrow \Omega_{bc}^{0}\gamma)$ 	&	0.145	 &	0.239	&	 $0.0039\pm0.0009$ 	&	 $0.579\pm0.014$ 	&	 $0.12\pm0.02$ 	&	-	&	-	 \\
    $\Gamma(\Omega_{cc}^{+*} \rightarrow \Omega_{cc}^{+}\gamma)$ 	&	1.37	 &	0.949	&	 $0.939\pm0.042$	&	 $1.973\pm0.029$ 	&	 $2.11\pm0.11$ 	&	0.82	&	 -	 \\
    \hline
    \end{tabular}
   \begin{center} 
\end{table}
\section{Semileptonic transition of doubly heavy baryons}\label{sec:4}
\textbf{\subsection{Form factors and Isgur-Wise function}}

One of the important topics in examining the features of DHBs is their weak decay rates. The study of semileptonic decays of heavy hadrons allows the determination of the Cabibbo-Kobayashi-Maskawa (CKM) matrix elements. Other properties of semileptonic decays, such as the momentum dependence of the transition form factors and exclusive decay rates are critical to our knowledge of heavy hadron structures.
\begin{figure}[h]
    \centering
    \includegraphics{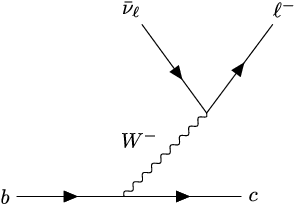}
    \caption{Feynman diagram for $b \rightarrow c$ semileptonic transition}
    \label{fig:1}
\end{figure}

The Feynman diagram of $b \rightarrow c$ transition is shown in Fig. \ref{fig:1}. Our focus is to study the semileptonic decays for $b \rightarrow c$ heavy-to-heavy transitions of the ground-state of DHBs. The differential decay width is given as \cite{Albertus2008}

\begin{eqnarray}\label{eq:20}
d\Gamma = 8 |V_{cb}|^2 m_{B'} G_F^2 \frac{d^3p'}{(2\pi)^3 2E'_{B'}} \frac{d^3k}{(2\pi)^3 2E_{\bar{\nu}_l}} \frac{d^3k'}{(2\pi)^3 2E'_l} \nonumber \\
 (2\pi)^4 \delta^4(p - p' - k - k') \mathcal{L}^{\alpha\beta}(k, k') \mathcal{H}_{\alpha\beta}(p, p')
\end{eqnarray}
where, $|V_{cb}|$ is the CKM matrix element. $m_{B'}$ is the mass of the final baryon, $G_F$ is the Fermi decay constant and $p$, $p'$, $k$ and $k'$ are the four-momenta of the initial baryon, final baryon, final anti-neutrino and final lepton, respectively. $\mathcal{L}$ and $\mathcal{H}$ are leptonic and hadron tensors, respectively, and are given as
\begin{eqnarray}\label{eq:21}
\mathcal{L^{\mu\sigma}}(k, k') = k'^\mu k^\sigma + k'^\sigma k^\mu - g^{\mu\sigma} k \cdot k' + i \epsilon^{\mu\sigma\alpha\beta} k'_\alpha k_\beta
\end{eqnarray}

\begin{eqnarray}\label{eq:22}
\mathcal{H}^{\mu\sigma}(p, p') = \frac{1}{2} \sum_{s,s'} \langle B', s'\vec{p}' \vert \bar{\Psi}^c(0) \gamma_\mu (I - \gamma_5) \Psi^b(0) \vert B, s \vec{p} \rangle \nonumber \\
\langle B', s' \vec{p}' \vert \bar{\Psi}^c(0) \gamma_\sigma (I - \gamma_5) \Psi^b(0) \vert B, s \vec{p} \rangle^*
\end{eqnarray}
$\vert B, s \vec{p}\rangle$ and $\vert B', s' \vec{p}'\rangle$ are the initial and final baryons with momenta $\vec{p}$ and third component of spin $s$. The baryon states are normalized as $\langle s \vec{p} \vert s' \vec{p}'\rangle = (2\pi)^3\frac{E(\vec{p})}{m}\\ \delta_{ss'}\delta^3(\vec{p}-\vec{p}')$.
The hadron matrix elements can be parameterized in terms of six form factors as
\begin{eqnarray}\label{eq:23}
\langle B', s' \vec{p}' \vert \bar{\Psi}^c(0) \gamma_\mu (I - \gamma_5) \Psi^b(0) \vert B, s \vec{p} \rangle \nonumber \\
 = \bar{u}^{B'}_{s'}(\vec{p}') \{\gamma_\mu(F_1(\omega)-\gamma_5G_1(\omega))+v_\mu(F_2(\omega)-\gamma_5G_2(\omega) \nonumber \\
 +v'_\mu (F_3(\omega)-\gamma_5G_3(\omega))\}u^B_s(\vec{p})
\end{eqnarray}
$\bar{u}^{B'}$ and $u^{B}$ are dimensionless Dirac spinors, normalised as $\bar{u}u=1$. $v_\mu=\frac{p_\mu}{m_B}$ and $v'_\mu=\frac{p'_\mu}{m_{B'}}$ are the four velocities of the initial $B$ and final $B'$ baryons, respectively.

    At zero recoil point, that is, $\omega$ = 1, $bb \rightarrow bc$ and $bc \rightarrow cc$ transitions become identical. The transversely polarized differential decay rate ($\Gamma_T$) and longitudinally polarized differential decay rate ($\Gamma_L$) neglecting the lepton masses, are given by,
\begin{eqnarray}\label{eq:24}
\frac{d\Gamma_T}{d\omega} = \frac{G_F^2 |V_{cb}|^2 m_{B'}^3 }{12\pi^3}  q^2 \sqrt{\omega^2 -1} \{(\omega-1)|F_1(\omega)|^2+(\omega+1)|G_1(\omega)|^2\}
\end{eqnarray}
\begin{eqnarray}\label{eq:25}
\frac{d\Gamma_L}{d\omega} = \frac{G_F^2 |V_{cb}|^2 m_{B'}^3 }{24\pi^3} \sqrt{\omega^2-1} \{(\omega-1)|\mathcal{F}^V(\omega)|^2+(\omega+1)|\mathcal{F}^A(\omega)|^2\}
\end{eqnarray}
\begin{eqnarray}\label{eq:26}
\mathcal{F}^{V,A}(\omega)=[(m_B\pm m_{B'})F_1^{V,A}(\omega)\nonumber \\
+(1\pm\omega)(m_{B'}F_2^{V,A}(\omega)+m_BF_3^{V,A}(\omega))]
\end{eqnarray}

$F_j^V \equiv F_j(\omega)$, $F_j^A \equiv G_j(\omega)$, $j$ = 1,2,3.
    In the heavy quark limit and close to zero recoil, the transition form factors are reduced to one, which is represented by the Isgur-Wise function $\eta$ \cite{Faessler2009}. Which is the function of the kinetic parameter $\omega$.

\begin{equation}\label{eq:27}
F_1(\omega)=G_1(\omega)=\eta(\omega)
\end{equation}
\begin{equation}\label{eq:28}
F_2(\omega) = F_3(\omega) = G_2(\omega) = G_3(\omega) = 0
\end{equation}

The Isgur-Wise function $\eta$ depends on $\omega$ which can be expressed as \cite{Albertus2005}

\begin{equation}\label{eq:29}
\eta(\omega) = \exp\left(-3(\omega - 1)\frac{m_{cc}^2}{\Lambda_B^2}\right)
\end{equation}

where $\omega$ = $v$ $\cdot$ $v'$ and $v$, $v'$ are the four velocities of the initial and final states of the DHBs, respectively. $\Lambda_B$ is the size parameter that varies in range 2.5 $\leq$ $\Lambda_B$ $\leq$ 3.5 GeV \cite{Faessler2009}.
The Isgur-Wise function can be calculated using Taylor's series expansion at the zero recoil point, $(\eta(\omega)\vert_{\omega=1} = 1)$ as $\eta(\omega) = 1-\rho^2(1-\omega)+c(1-\omega)^2+\ldots$ where, $\rho^2$ is the magnitude of the slope and $c$ is the curvature (convexity parameter) of the Isgur-Wise function $\eta(\omega)$ at $\omega=1$ can be written as
\begin{equation}\label{eq:30}
\rho^2 = -\frac{d\eta(\omega)}{d\omega}\vert_{\omega=1};   c=\frac{d^2\eta(\omega)}{d\omega^2}\vert_{\omega=1}
\end{equation}
\begin{equation}\label{eq:31}
\rho^2 = \frac{3m_{cc}^2}{\Lambda_b^2}; c=\frac{9m_{cc}^4}{2\Lambda_b^4}
\end{equation}

\textbf{\subsection{Differential decay widths}}

The differential decay rates from Eq. (\ref{eq:26})
\begin{equation}\label{eq:32}
\frac{d\Gamma_T}{d\omega} = \frac{G_F^2 |V_{cb}|^2 m_{B'}^3 }{6\pi^3}  q^2 \omega \sqrt{\omega^2 -1} \eta^2(\omega)
\end{equation}

\begin{eqnarray}\label{eq:33}
\frac{d\Gamma_L}{d\omega} = \frac{G_F^2 |V_{cb}|^2 m_{B'}^3 }{24\pi^3} \times[(\omega - 1)(m_B + m_{B'})^2 \nonumber \\
+ (\omega + 1)(m_B - m_{B'})^2]\eta^2(\omega)
\end{eqnarray}

 where, $q^2$ is the squared four-momentum transfer between the heavy baryons given as $q^2$ = $(p - p')^2$ = $m_{B}^2$+$m_{B'}^2$$-$$2m_{B}$$m_{B'}\omega$, where $m_{B}$ and $m_{B'}$ are the masses of the initial and final baryons, respectively. We have taken $|V_{cb}|$ = 0.042. The total differential decay rate is given by

\begin{equation}\label{eq:34}
\frac{d\Gamma}{d\omega} = \frac{d\Gamma_T}{d\omega} + \frac{d\Gamma_L}{d\omega}
\end{equation}

The total decay width is calculated by integrating the total differential decay rate from 1 to $\omega_{max}$ maximal recoil ($q^2 = 0$). The obtained values of $\omega_{max}$ for different transitions are listed in Table \ref{tab:table11}.
\begin{equation}\label{eq:35}
\Gamma = \int_{1}^{\omega_{max}} \frac{d\Gamma}{d\omega} \, d\omega
\end{equation}
\begin{equation}\label{eq:36}
\omega_{max} = \frac{m_B^2 + m_{B'}^2}{2m_B m_{B'}}
\end{equation}

\begin{equation}\label{eq:37}
Br = \Gamma\times\tau
\end{equation}


\begin{figure}
\centering
\includegraphics[scale=0.33]{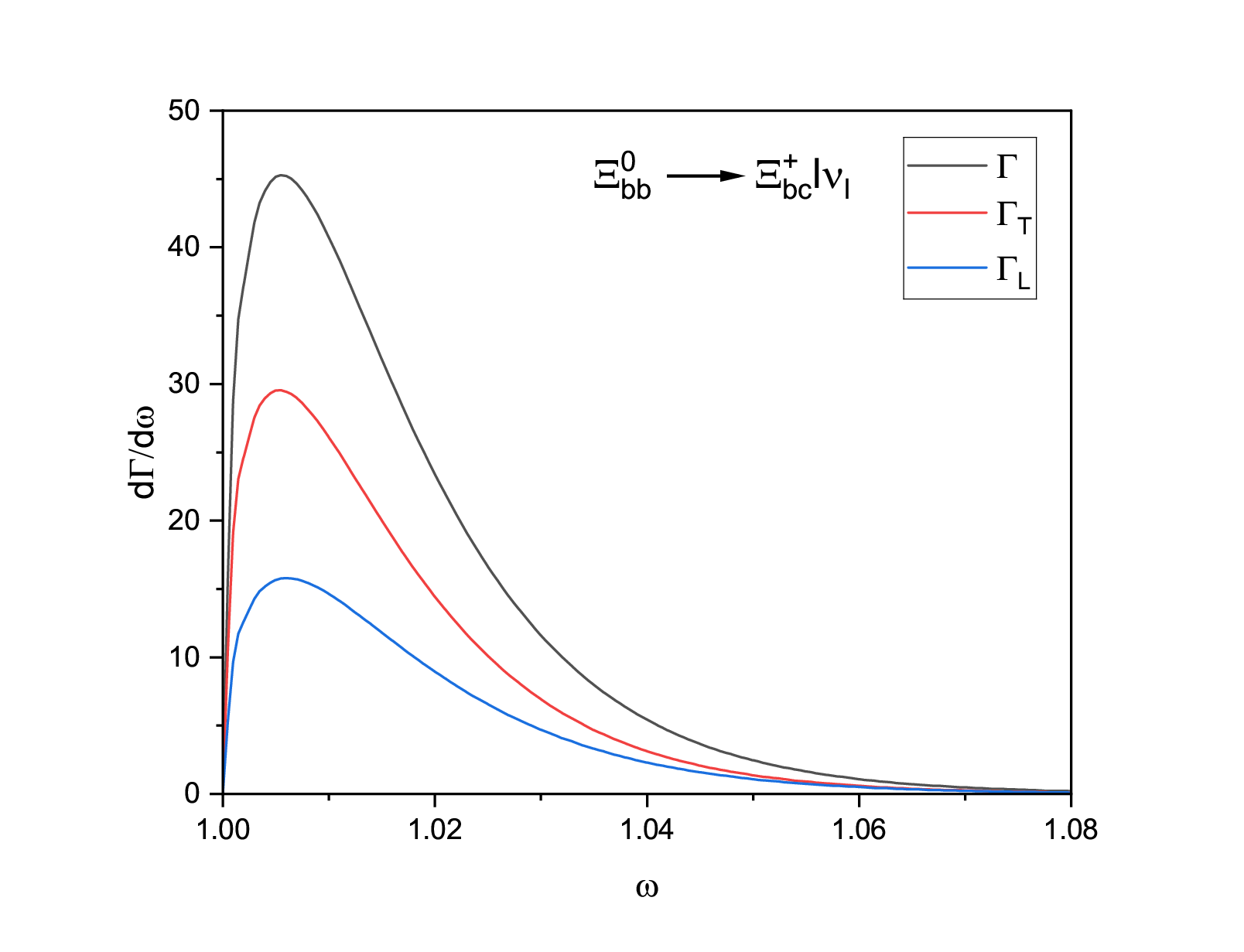}
\caption{\label{fig:2}Differential decay rates for $\Xi_{bb}^0 \rightarrow \Xi_{bc}^+ \ell\bar{\nu}_\ell$ transition}
\end{figure}

\begin{figure}
\centering
\includegraphics[scale=0.33]{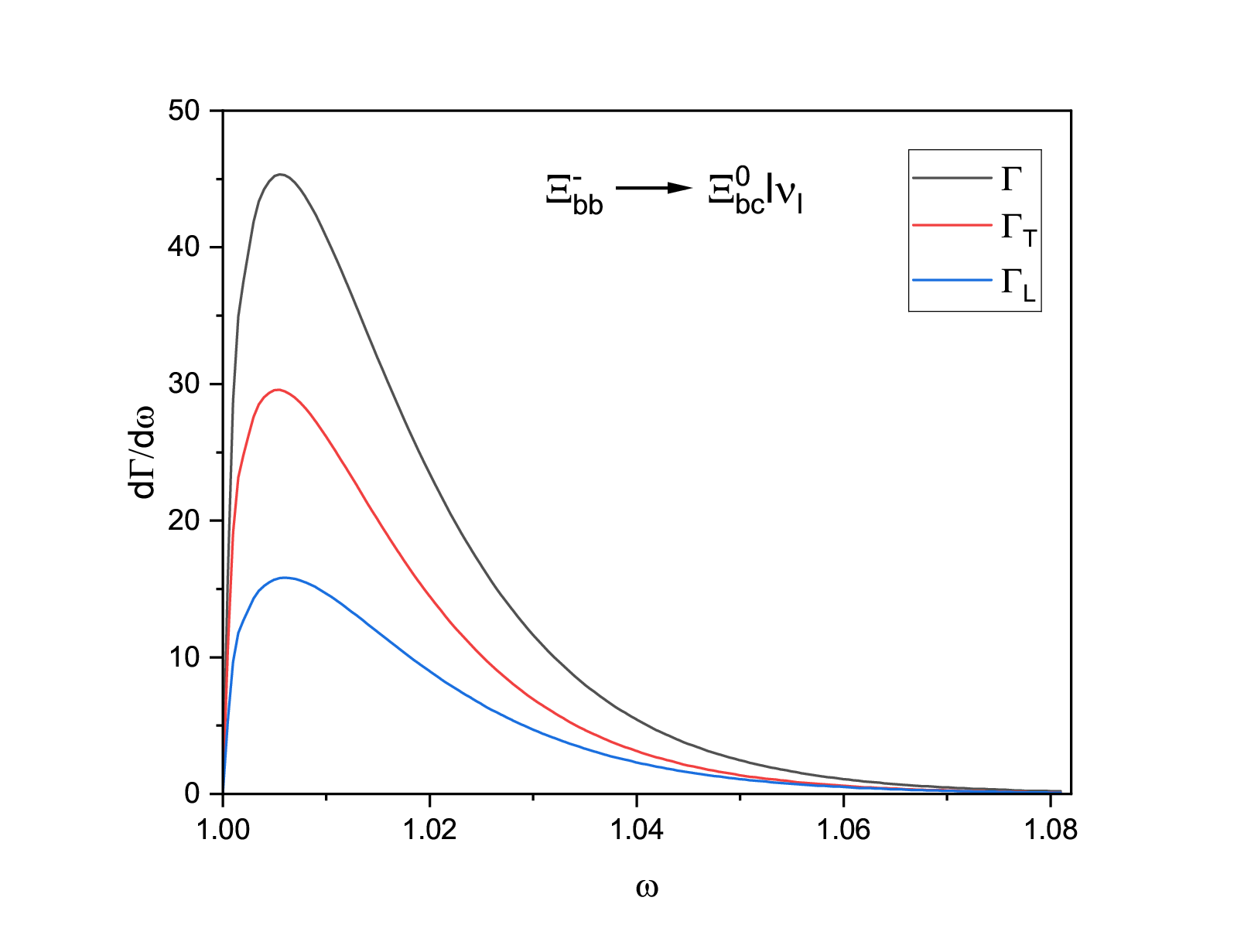}
\caption{\label{fig:3}Differential decay rates for $\Xi_{bb}^- \rightarrow \Xi_{bc}^0 \ell\bar{\nu}_\ell$ transition}
\end{figure}

\begin{figure}
\centering
\includegraphics[scale=0.33]{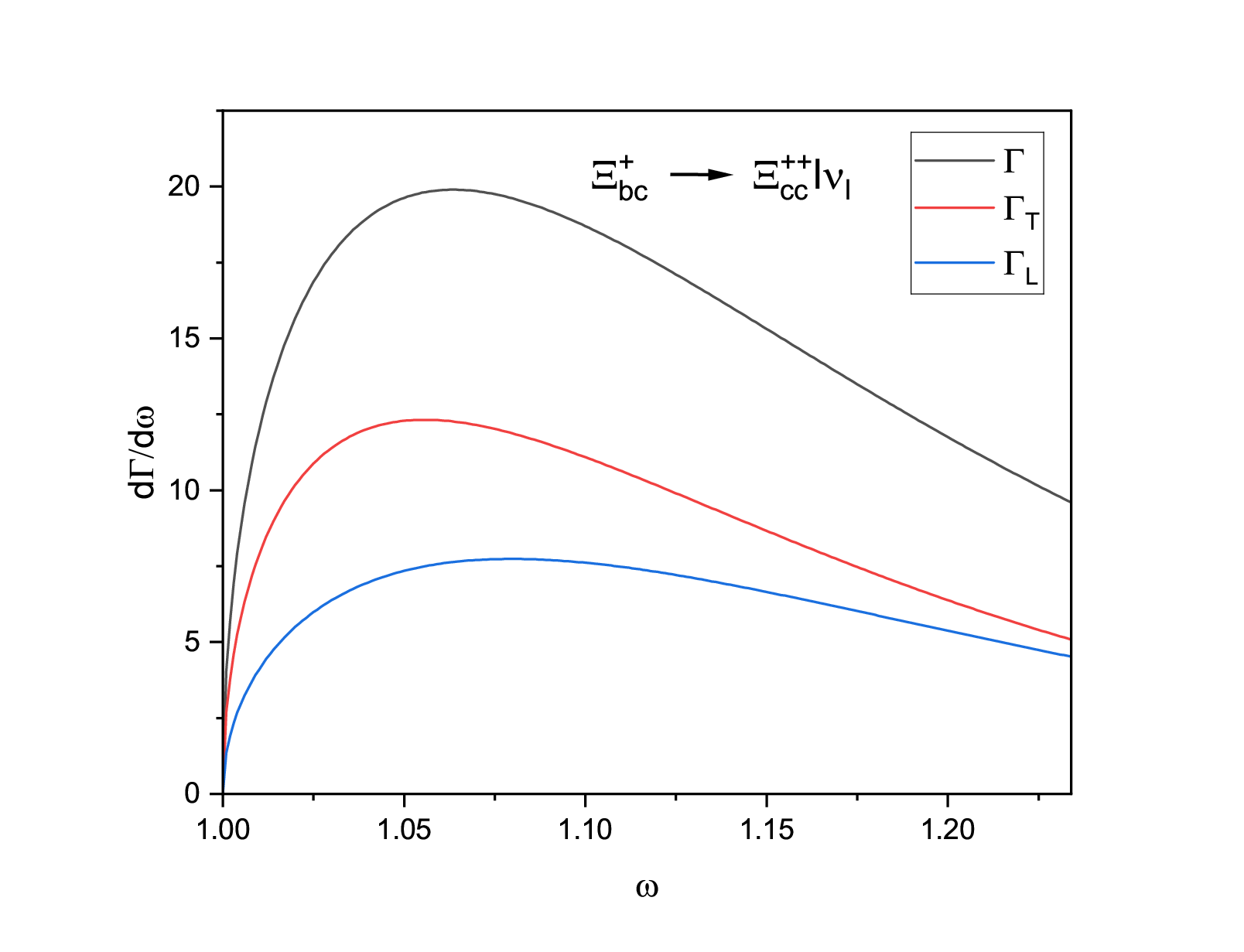}
\caption{\label{fig:4}Differential decay rates for $\Xi_{bc}^+ \rightarrow \Xi_{cc}^{++} \ell\bar{\nu}_\ell$ transition}
\end{figure}

\begin{figure}
\centering
\includegraphics[scale=0.33]{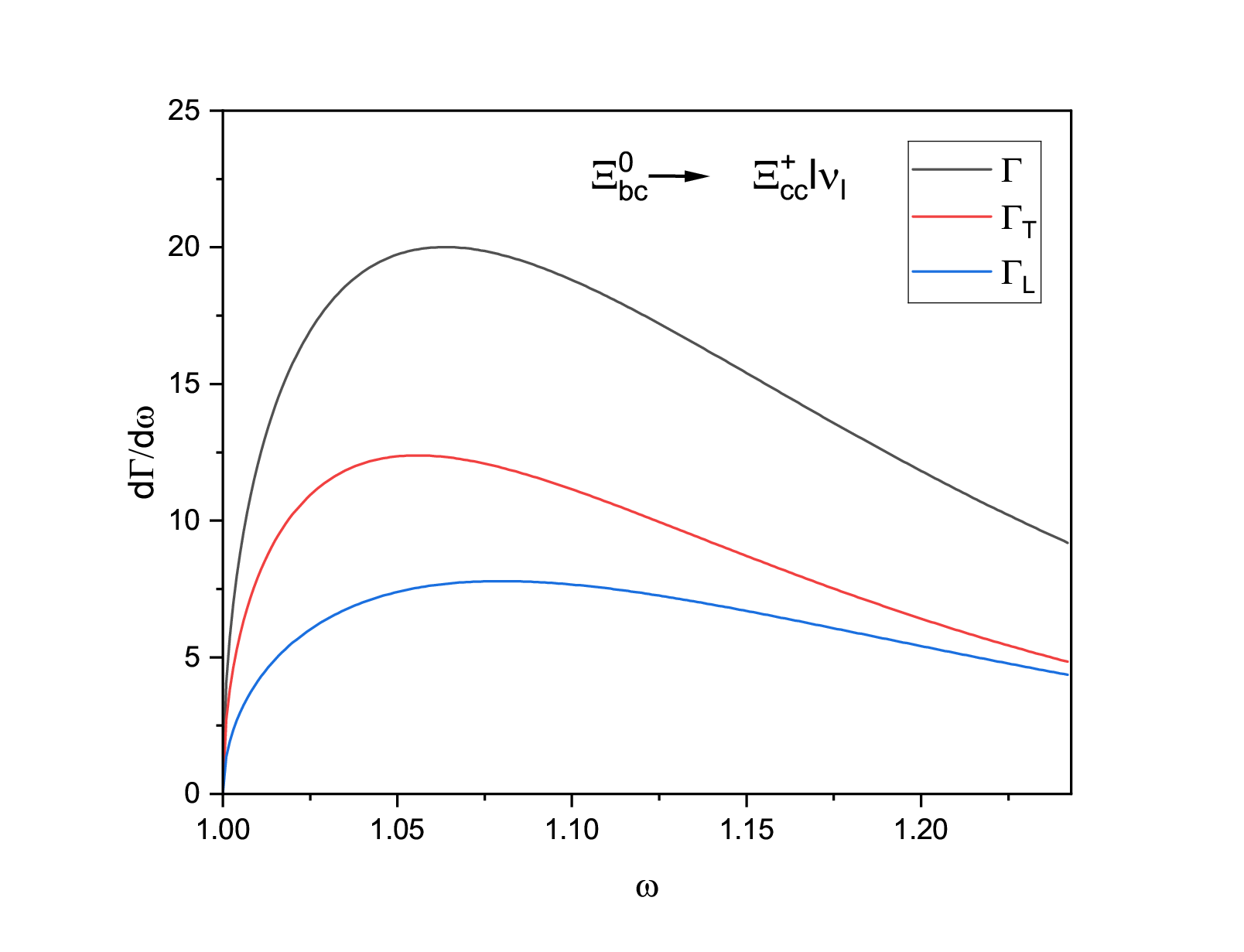}
\caption{\label{fig:5}Differential decay rates for $\Xi_{bc}^0 \rightarrow \Xi_{cc}^+ \ell\bar{\nu}_\ell$ transition}
\end{figure}

\begin{figure}
\centering
\includegraphics[scale=0.33]{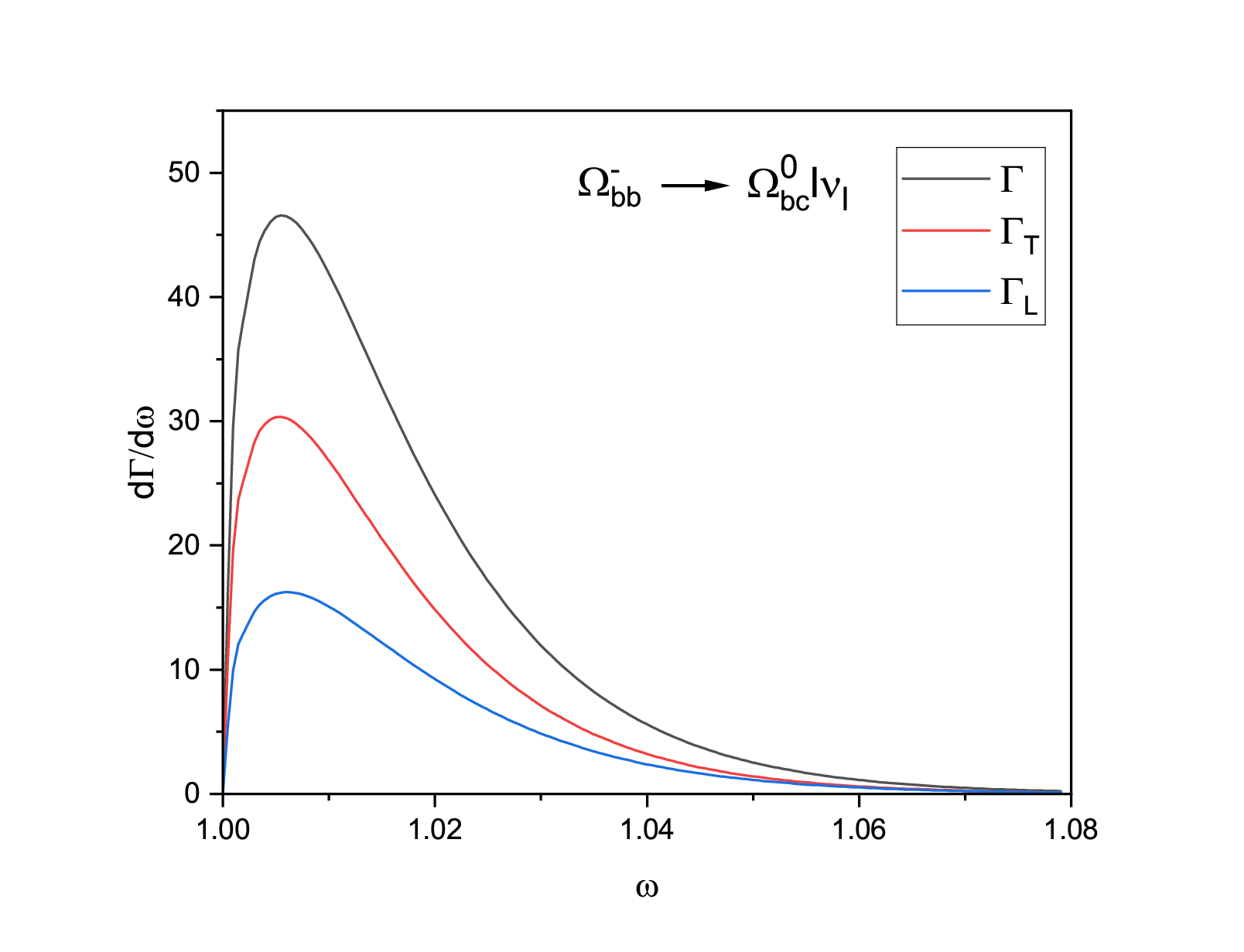}
\caption{\label{fig:6}Differential decay rates for $\Omega_{bb}^- \rightarrow \Omega_{bc}^0 \ell\bar{\nu}_\ell$ transition}
\end{figure}

\begin{figure}
\centering
\includegraphics[scale=0.33]{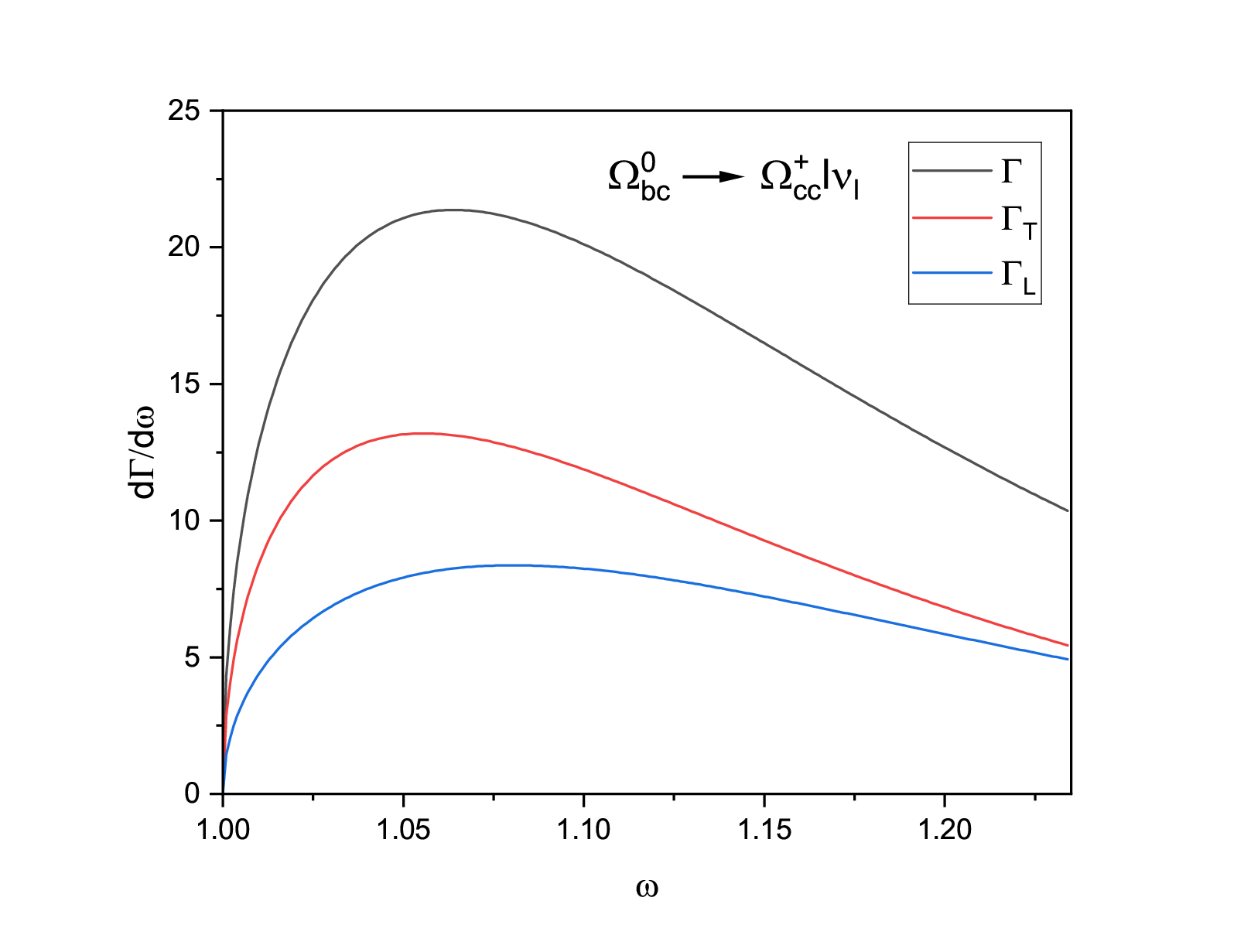}
\caption{\label{fig:7}Differential decay rates for $\Omega_{bc}^0 \rightarrow \Omega_{cc}^+ \ell\bar{\nu}_\ell$ transition}
\end{figure}


\begin{figure}
\centering
\includegraphics[scale=0.33]{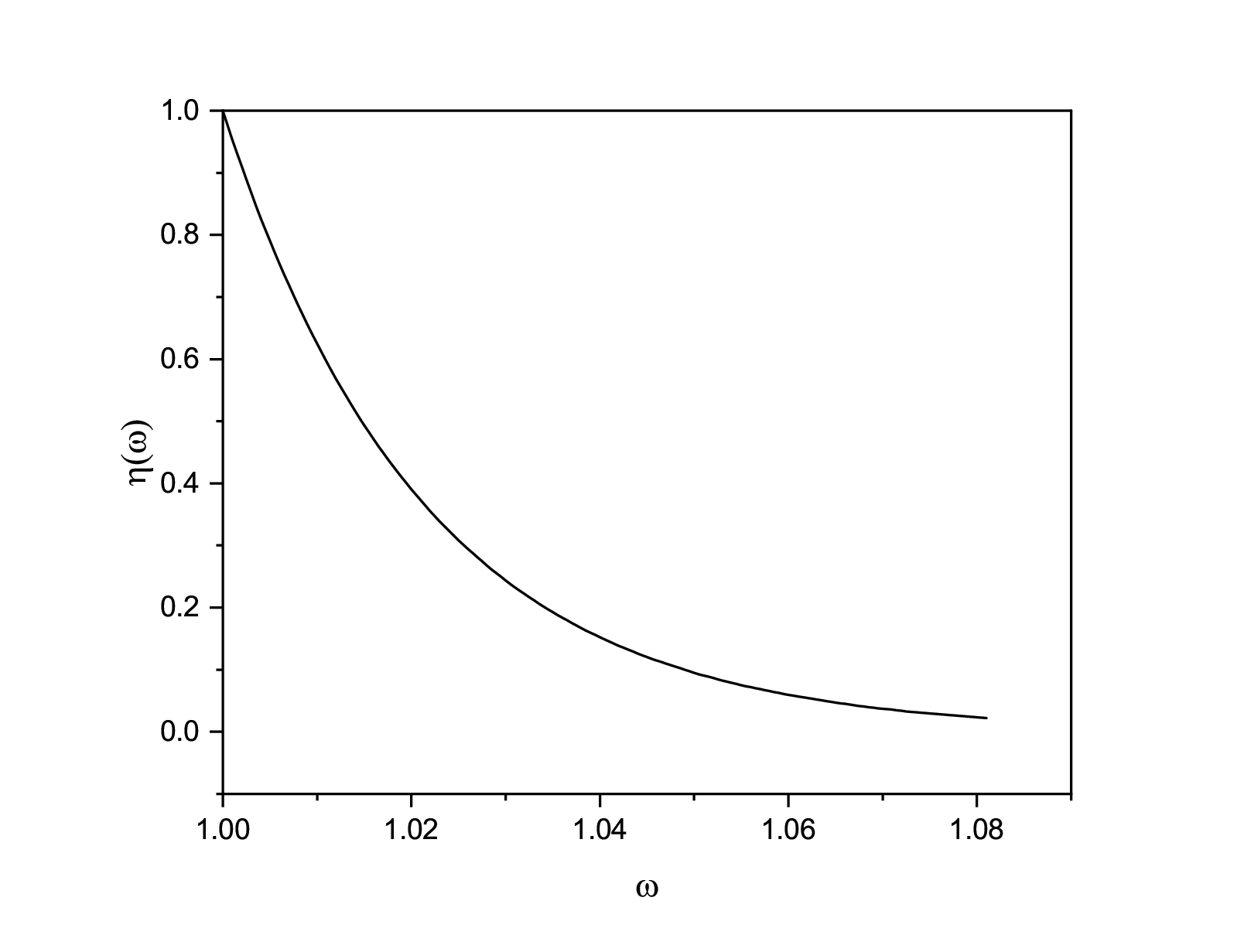}
\caption{\label{fig:8}Isgur-Wise function $\eta(\omega)$ for $\Xi_{bb}^0 \rightarrow \Xi_{bc}^+ \ell\bar{\nu}_\ell$ transition }
\end{figure}

\begin{figure}
\centering
\includegraphics[scale=0.33]{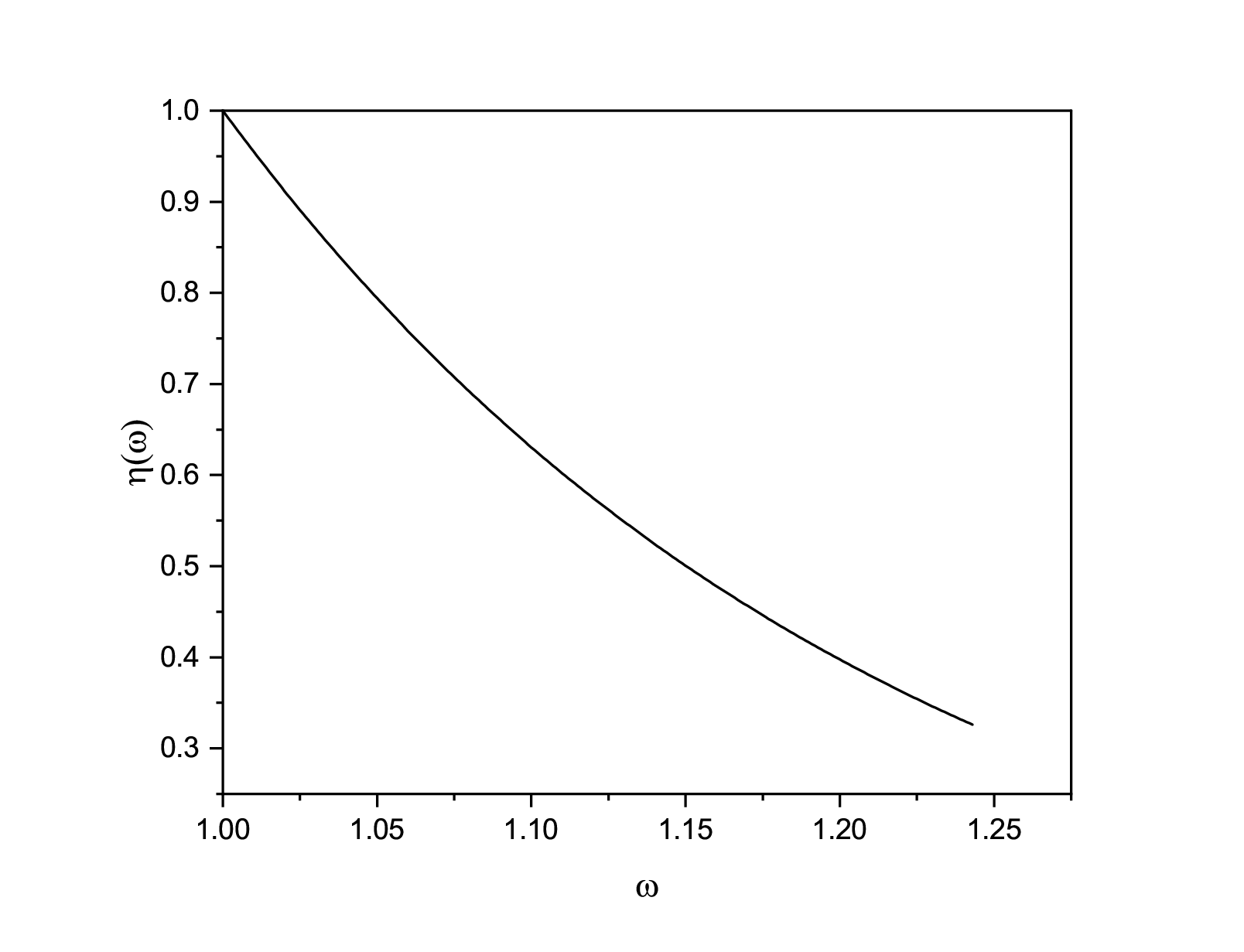}
\caption{\label{fig:9}Isgur-Wise function $\eta(\omega)$ for $\Xi_{bc}^+ \rightarrow \Xi_{cc}^{++} \ell\bar{\nu}_\ell$ transition}
\end{figure}

    The branching ratio of DHBs can be calculated using Eq. (\ref{eq:37}), where $\tau$ is the lifetime of the initial baryon.\\

\section{Results and discussions}\label{sec:5}
    In this study, we focused on the heavy-to-heavy semileptonic and radiative decay widths of ground-state DHBs. We also computed the ground-state masses, magnetic moments, and transition magnetic moments of DHBs. The Hypercentral Constituent Quark Model (HCQM) simplifies the three-body problem of DHBs using hyperspherical coordinates by solving the Schr\"{o}dinger equation with a hypercoulomb plus linear potential. In this work, we have not included the $\Xi_{bc}'$ and $\Omega_{bc}'$ baryons. Although both baryons have been studied theoretically, they are yet to be observed experimentally. These baryons are composed of an antisymmetric spin-0 which is generally calculated in Quark-Diquark model. They are composed of a light quark and a heavy diquark in an antisymmetric spin-0 state. Such a configuration is generally treated more suitably within the Quark-Diquark approximation see Ref. \cite{Zhang2008,Tang2012}, where the baryon is approximated as a two-body system-unlike the fully three-body treatment employed in the present HCQM. \\

  $\bullet$ \emph{Ground State Masses of Doubly Heavy Baryons}\\

    \begin{enumerate}
    \item  We have calculated the ground-state masses of all the DHBs using the parameters shown in Table \ref{tab:Table2}. We set the same parameters for all the DHBs using which we evaluated various properties of DHBs. The calculated masses of the ground-state DHBs are listed in Table \ref{tab:Table3}. The mass difference between the up quark and down quark is neglected in all other theoretical predictions. In the present work, we have considered different quark masses for up and down quarks as $m_u$ = 0.33 GeV and $m_d$ = 0.35 GeV (see Table \ref{tab:Table2}). \\
    \item Our calculated masses of DHBs are in agreement with other theoretical predictions, especially with Ref. \cite{Ebert2005}. The values obtained in Ref. \cite{Tousi2024} are smaller than our calculated masses. This may be due to the incorporation of ten mass dimensions non-perturbative operators in QCD sum rule formalism, which assumed the masses of the $u$ and $d$ quarks to be zero. \\
    \item The percentage errors for the ground-state masses are calculated as
    \begin{equation}\label{eq:38}
    \% Error = \vert\frac{M_{HCQM}-M_{Lattice}}{M_{Lattice}}\vert\times100
    \end{equation}
    The calculated percentage errors for the ground-state masses are listed in Table \ref{tab:Table3}. Due to the lack of experimental data, we have used the Lattice QCD masses to calculate the error. The Lattice masses are taken from Ref. \cite{Brown2014} and \cite{Mathur2018}. As seen in Table \ref{tab:Table3}, the percentage errors for doubly bottom baryons and bottom-charm baryons range from 0.35$\%$ to 1.27$\%$ whereas the errors for doubly charmed baryons are of the order of 5$\%$. Overall, the small percentage errors (mostly under 5$\%$) indicate good agreement between our calculated masses and Lattice predictions.\\
    \end{enumerate}

$\bullet$ \emph{Magnetic Moments and Transition Magnetic Moments}\\
\begin{enumerate}
    \item As shown in Table \ref{tab:table6}, the magnetic moments of the DHBs are almost matched with other models. The magnitude and sign of the magnetic moments provide insights into the size, structure and shape deformations of baryons. Here, the dominant contribution to the magnetic moments of the DHBs came from the light quark. The magnetic moment of $\Xi^{0*}_{bb}$ predicted in Ref. \cite{shah2021} has a negative value, whereas all other theoretical approaches, including ours, predict positive values. \cite{Liu2018} reports the lowest values for magnetic moments of $\Xi^{+}_{cc}$ = $0.392\pm0.013$ $\mu_N$ and $\Omega^{+}_{cc}$= $0.397\pm0.015$ $\mu_N$ calculated within covariant baryon chiral perturbation theory with the extended-on-mass-shell scheme up to the next-to-leading order. Discrepancies in the magnetic moment of $\Xi^{++}_{cc}$ baryon were obsereved. The signs of the magnetic moments are correctly determined. In some cases, different theoretical models have yielded quite different results for the magnetic moments, this may be due to the choice of wave functions and assumptions in different models.\\
    \item The transition magnetic moments of the DHBs are listed in Table \ref{tab:table7}. The expressions of the transition magnetic moments for DHBs are derived from the spin-flavour wave functions. As indicated in Table \ref{tab:table7}, there is good agreement between the computed transition magnetic moments and the other predictions except for Ref. \cite{Bernotas2013}, which has relatively low values. In Ref. \cite{Bernotas2013}, the framework of the modified bag model is adopted and the transition magnetic moments are obtained using the radii of lighter baryons under the assumption of light quarks $u$ and $d$ to be massless. The change in the sign of $\Xi_{bc}^{+*} \rightarrow \Xi_{bc}^{+}$, $\Xi_{bc}^{0*} \rightarrow \Xi_{bc}^{0}$ and $\Omega_{bc}^{0*}\rightarrow \Omega_{bc}^{0}$ transitions arises from a positive shift due to hyperfine mixing effects. LQCD \cite{Bahtiyar2018} reports the lowest values than all the approaches of transition magnetic moments for $\Xi_{cc}^{++*} \rightarrow\Xi_{cc}^{++}$ = $-0.772$, $\Xi_{cc}^{+*}  \rightarrow\Xi_{cc}^{+}$ = 0.906 and $\Omega_{cc}^{+*} \rightarrow \Omega_{cc}^{+}$ = 0.882. \\
\end{enumerate}
$\bullet$ \emph{Radiative decay widths}\\
\begin{enumerate}
    \item The radiative decay widths depend on the magnitude of the transition magnetic moment and photon momentum $k$. Comparing the radiative decay width with other models, we found that different approaches led to different results, as shown in Table \ref{tab:table8}. The results for the radiative decay widths are sensitive to the mass difference of initial and final doubly heavy baryon. We can see that the radiative decay width is relatively large for the $\Xi_{cc}^{++*} \rightarrow\Xi_{cc}^{++}\gamma$ and $\Xi_{cc}^{+*} \rightarrow\Xi_{cc}^{+}\gamma$ in the relativistic three quark model \cite{Branz2010} while comparing with others. The lattice calculations \cite{Bahtiyar2018} reports comparatively lower values for the radiative decays $\Xi_{cc}^{++*} \rightarrow\Xi_{cc}^{++}\gamma$ = 0.0518 keV, $\Xi_{cc}^{+*} \rightarrow\Xi_{cc}^{+}\gamma$ = 0.0648 keV and $\Gamma(\Omega_{cc}^{+*} \rightarrow \Omega_{cc}^{+}\gamma)$ = 0.0565 keV which is due to the decrease in the kinematic factors and uncontrolled systematic errors. LCQR \cite{Aliev2023} reports higest values for $\Xi_{bb}^{0,-}$ baryon transitions. Our computed radiative decay width for $\Omega_{bb}^{-*} \rightarrow \Omega_{bb}^{-}\gamma$ transition is relatively lower than all other predictions. The discrepancies in radiative decay widths can be due to the photon momenta $k$, which depends on the $J^P= \frac{3}{2}^+$ and $\frac{1}{2}^+$ masses of baryons. Hence, the results for the radiative decay widths differs with the different choices of the messes of the baryons. Future experimental efforts on DHBs can resolve these discrepancies among the different model predictions of magnetic moments and radiative decay widths.\\
\end{enumerate}
$\bullet$ \emph{Isgur-Wise functions and Semileptonic Decay widths}\\
\begin{enumerate}
     \item The semileptonic decays of DHBs are useful tools for extracting the CKM matrix elements. To calculate the semileptonic decay rate, we have considered $m_{bb}$ = 2$m_b$ = 9.9 GeV and for $bc \rightarrow cc$ transition, $m_{cc}$ = 2$m_c$ = 3.1 GeV in the Eq. (\ref{eq:29}). We have considered the size parameter $\Lambda_B$ = 2.5 GeV \cite{Rahmani2020,Ivanov1997}. The calculated semileptonic decay rates of the baryons are listed and compared with those of other models in Table \ref{tab:table9}. The present results for the semileptonic decay width of DHBs are close to the results predicted in Ref. \cite{Ghalenovi2023}. The variations observed in the results arise from differences in the assumptions regarding the diquark structure, quark mass values and Isgur-Wise function parameterisation. Different forms for the Isgur-Wise function may result in different decay widths. However, the order of decay width is not changed. Additionally, the results are insensitive to the size parameter $\Lambda_B$, a smaller value of $\Lambda_B$ gives smaller decay widths and vice versa.\\
     \item The total differential decay rate ($\frac{d\Gamma}{d\omega}$) can be written as the summation of the transverse differential decay rate ($\frac{d\Gamma_T}{d\omega}$) and longitudinal decay rate ($\frac{d\Gamma_L}{d\omega}$) as indicated in Eq.(\ref{eq:34}). It is found that the contribution from the transverse decay ($\Gamma_T$) is relatively higher compared to the longitudinal decay ($\Gamma_L$), as shown in Table \ref{tab:table10}. Approximately $60\%$ of the contribution comes from $\Gamma_T$, whereas $\Gamma_L$ contributes around $40\%$.\\
    \item The behaviour of the variation of the Isgur-Wise function with respect to $\omega$ is shown in Fig. \ref{fig:8} and \ref{fig:9}. The plots for $\Xi_{bb}^-$ and $\Omega_{bb}^-$ are (not shown) similar to Fig. \ref{fig:8} while for $\Xi_{bc}^0$ and $\Omega_{bc}^0$  are (not shown) similar to Fig. \ref{fig:9}. It can be seen that the $bb \rightarrow bc$ transition decays faster as $\omega$ increases, because of the larger $m_{bb}$ = 9.9 GeV. For the $bc \rightarrow cc$ transition, the Isgur-Wise function decreases gradually with increasing $\omega$ due to $m_{cc}$ = 3.1 GeV. The curve for $bb \rightarrow bc$ is steeper, with the Isgur-Wise function approaching zero much faster. The plots show how the transition behaves differently based on the underlying mass parameters. The slope and curvature (convexity parameter) of the Isgur-Wise function are constant, with parameter $m_{cc}=3.1$ for bottom-charmed baryons and $m_{bb}=9.9$ for doubly bottom baryons.\\
    \item The behaviour of the predicted differential decay rates for the semileptonic decay of DHBs with $\omega$ are shown in Figure \ref{fig:2}--\ref{fig:7}. The peak value of differential decay rate ($\frac{d\Gamma}{d\omega}$) for $\Xi_{bb}$ and $\Omega_{bb}$ baryons is found at $\omega\approx1.01$ while the peak value for $\Xi_{bc}$ and $\Omega_{bc}$ baryons is found at $\omega\approx1.06$. The $\frac{d\Gamma}{d\omega}$ of $\Xi_{bb}$ and $\Omega_{bb}$ baryons gets saturated around $\omega\approx1.06$ while $\Xi_{bc}$ and $\Omega_{bc}$ baryons are at peak value for $\omega\approx1.06$.\\
\end{enumerate}
$\bullet$ \emph{Branching Ratios of Doubly Heavy Baryons}\\
\begin{enumerate}
     \item We calculate the branching ratios for DHBs using Eq. \ref{eq:37} (see Table \ref{tab:table12}). The lifetimes of baryons have been studied in Ref. \cite{Berezhnoy2018,Kiselev2002,Likhoded2018,Cheng2019}. To calculate the branching ratios, we have considered $\tau_{\Xi_{bb}^0}$ = 0.52$\times10^{-12}$ s, $\tau_{\Xi_{bb}^-}$ = 0.53 $\times10^{-12}$ s, $\tau_{\Xi_{bc}^+}$ = 0.24$\times10^{-12}$ s, $\tau_{\Xi_{bc}^0}$ = 0.22$\times10^{-12}$ s, $\tau_{\Omega_{bb}^-}$ = 0.53 $\times10^{-12}$ s, $\tau_{\Omega_{bc}^0}$ = 0.18$\times10^{-12}$ s as given in Ref. \cite{Berezhnoy2018}. To ensure a fair comparison of our branching ratio results with other theoretical predictions, we have computed the branching ratio for all the models compared in Table \ref{tab:table9}, using their predicted semileptonic decay widths and lifetime values given in Ref. \cite{Berezhnoy2018}. Overall, our computed branching ratios are in good agreement with those of other theoretical models.\\
\end{enumerate}
\begin{table*}
    \centering
    \caption{\label{tab:table9}The semileptonic decay width of DHBs $\Gamma$ in $ 10^{-14} GeV$}
    \begin{tabular}{ccccccccc}
    \hline
    Decay & Our & \cite{Ebert2005} & \cite{Ghalenovi2022} & \cite{Ghalenovi2023} & \cite{Faessler2009} & \cite{Rahmani2020} & \cite{Albertus2008} & \cite{Wang2017}\\
    \hline
    $\Xi_{bb}^0 \rightarrow \Xi_{bc}^+\ell\bar{\nu}_\ell$ & 1.0526 & 3.26 & 1.75 & 0.98 & 0.8 & 0.49 & 1.92 & 3.30\\
    $\Xi_{bb}^- \rightarrow \Xi_{bc}^0\ell\bar{\nu}_\ell$ & 1.0539 & -&- &- & -&- & - &3.30\\
    $\Xi_{bc}^+ \rightarrow \Xi_{cc}^{++}\ell\bar{\nu}_\ell$ & 4.1456& 4.59 & 3.08 & 4.39 & 2.1 & 3.01 & 2.57 & 4.50\\
    $\Xi_{bc}^0 \rightarrow \Xi_{cc}^+\ell\bar{\nu}_\ell$ & 4.1589 &- &- & -&- &- & - & 4.50  \\
    $\Omega_{bb}^- \rightarrow \Omega_{bc}^0\ell\bar{\nu}_\ell$ & 1.0828 & 3.40 & 1.03 & 1.87 & 0.86 & 0.99 & 2.14 & 3.69\\
    $\Omega_{bc}^0 \rightarrow \Omega_{cc}^+\ell\bar{\nu}_\ell$ & 4.3336 & 4.95 & 3.32 & 4.7 & 1.88 & 3.28 & 2.59 & 3.94\\
    \hline
    \end{tabular}
\end{table*}

\begin{table}
    \centering
    \caption{\label{tab:table10}The transverse $\Gamma_T$ and longitudinal $\Gamma_L$ contributions to the width in $10^{-14} GeV$}
    \begin{tabular}{ccccc}
    \hline
    Decay & Our & Our  &\cite{Ghalenovi2023} & \cite{Ghalenovi2023}  \\
    & $\Gamma_T$ & $\Gamma_L$ & $\Gamma_T$ & $\Gamma_L$\\
    \hline
    $\Xi_{bb}^0	\rightarrow	\Xi_{bc}^+\ell\bar{\nu}_\ell$	&	0.659	&	 0.393	 &	 0.55	&	0.42\\
    $\Xi_{bb}^-	\rightarrow	\Xi_{bc}^0\ell\bar{\nu}_\ell$	&	0.660	&	 0.393	 &	-&-\\		
    $\Xi_{bc}^+	\rightarrow	\Xi_{cc}^{++}\ell\bar{\nu}_\ell$	&	2.437	&	 1.708	 &	 1.32	&	1.75\\
    $\Xi_{bc}^0	\rightarrow	\Xi_{cc}^+\ell\bar{\nu}_\ell$	&	2.445	&	 1.713	 &-&-	 \\		
    $\Omega_{bb}^-	\rightarrow	\Omega_{bc}^0\ell\bar{\nu}_\ell$	&	0.678	 &	 0.404	&	0.58	&	0.45\\
    $\Omega_{bc}^0	\rightarrow	\Omega_{cc}^+\ell\bar{\nu}_\ell$	&	2.543	 &	 1.790	&	 1.4	&	1.91\\
    \hline
    \end{tabular}
\end{table}

\begin{table}
 \centering
    \caption{\label{tab:table11}Obtained values of $\omega_{max}$ for $b \rightarrow c$ transitions}
    \begin{tabular}{ccc}
    \hline
    Transition &	our	&	\cite{Ghalenovi2022}\\
    \hline
    $\Xi_{bb}^0	\rightarrow	\Xi_{bc}^+\ell\bar{\nu}_\ell$	&	1.082	&	 1.07\\
    $\Xi_{bb}^-	\rightarrow	\Xi_{bc}^0\ell\bar{\nu}_\ell$	&	1.081	&	-\\
    $\Xi_{bc}^+	\rightarrow	\Xi_{cc}^{++}\ell\bar{\nu}_\ell$ &	1.244	&	 1.22\\
    $\Xi_{bc}^0	\rightarrow	\Xi_{cc}^+\ell\bar{\nu}_\ell$	&	1.243	&	-\\
    $\Omega_{bb}^- \rightarrow \Omega_{bc}^0\ell\bar{\nu}_\ell$	&	1.080	&	 1.07\\
    $\Omega_{bc}^0 \rightarrow \Omega_{cc}^+\ell\bar{\nu}_\ell$	&	1.233	&	 1.20\\
    \hline
    \end{tabular}
\end{table}

\begin{table*}
\centering
    \caption{\label{tab:table12}Branching Ratio in ($\%$), calculated for all models using lifetimes given in Ref. \cite{Berezhnoy2018}.}
    \begin{tabular}{cccccccccc}
    \hline
    Transition & Our & \cite{Ebert2005} & \cite{Ghalenovi2022} & \cite{Ghalenovi2023} & \cite{Faessler2009} & \cite{Rahmani2020} & \cite{Albertus2008}  & \cite{Wang2017}  \\
    \hline
    $\Xi_{bb}^0 \rightarrow \Xi_{bc}^+l\bar{\nu}_l$	&	0.831	&	1.288	&	 1.383	 &	1.489	&	0.632	&	0.742	&	1.517	&	2.607	\\
    $\Xi_{bb}^- \rightarrow \Xi_{bc}^0l\bar{\nu}_l$	&	0.849	&-	& - &-  & -  &  - & -  & -    \\													
    $\Xi_{bc}^+ \rightarrow \Xi_{cc}^{++}l\bar{\nu}_l$ &	1.512	&	0.839	 &	 1.601	&	1.123	&	0.766	&	1.098	&	0.937	&	1.641		 \\
    $\Xi_{bc}^0 \rightarrow \Xi_{cc}^+l\bar{\nu}_l$	&	1.390	&	-&-	& - &  - & -  & -  &  -   \\													
    $\Omega_{bb}^- \rightarrow \Omega_{bc}^0l\bar{\nu}_l$ &	0.872	&	1.369	 &	 1.506	&	0.829	&	0.692	&	0.797	&	1.723	&	2.971		 \\
    $\Omega_{bc}^0 \rightarrow \Omega_{cc}^+l\bar{\nu}_l$ &	1.118	&	0.678	 &	 1.285	&	0.908	&	0.514	&	0.897	&	0.708	&	1.077	 \\
    \hline
    \end{tabular}
\end{table*}

\section{Conclusions}\label{sec:6}
 We have calculated the static and dynamic properties of DHBs in the framework of Hypercentral Constituent Quark Model(HCQM). The ground-state masses are calculated by solving the six-dimensional Schr\"{o}dinger equation. The magnetic moments of DHBs are computed using the spin-flavour wave functions of the constituent quarks and their effective masses within the baryon. We have calculated the radiative $M1$ decay width from the obtained transition magnetic moment for the $\frac{3}{2}^+ \rightarrow \frac{1}{2}^+$ transitions. The semileptonic decay rates for DHBs are calculated after obtaining the Isgur-Wise function. Additionally, the transverse and longitudinal components of the semileptonic decay widths are calculated. Finally, the branching ratios are obtained from the computed semileptonic decay rates.\\

\textbf{Data availability statement} This is a theoretical work and it does not contain any data.


\end{document}